\documentclass{ptapap}
\usepackage{color}
\definecolor{todo}{rgb}{0.8,0.2,0.3}
\author{Andrzej Pigulski}[IAUWr]
\author{the BRITE Team}
\affil[IAUWr]{Instytut Astronomiczny, Uniwersytet Wroc{\l}awski, Kopernika 11, 51-622 Wroc{\l}aw, Poland}
\title{Statistical Overview of BRITE Targets Observed So Far}
\begin{document}
\maketitle
\begin{abstract}
We characterize BRITE data obtained between 2013 and 2017 in the first 21 BRITE observing fields. Then, we overview the sample of 426 stars observed so far by the BRITE satellites. The review shows that BRITEs provide unique and precise space photometry, which allows to obtain outstanding scientific results in many areas of stellar astrophysics.
\end{abstract}
\section{Introduction}
The BRIght Target Explorer (BRITE) nano-satellites, launched between March 2013 and August 2014, already have provided space photometry of bright stars for more than three years.  Up to the time of writing, the data for 426 bright stars in 21 fields (some re-observed) were released to users. The analysis of the two-band BRITE photometry have already led to many interesting discoveries (see, for example, the highlights presented by D.\,Baade, these proceedings, and Sect.~\ref{science}). It is therefore a good time to overview the statistics of the observed targets and the data obtained with this ongoing space mission.

The BRITE mission consists of a constellation of five nano-satellites\footnote{The names of the BRITE satellites are abbreviated to: BRITE-Austria (BAb), UniBRITE (UBr), BRITE-Lem (BLb), BRITE-Toronto (BTr), and BRITE-Heweliusz (BHr).} on low orbits, of which two are equipped in blue filters and the remaining three in red filters. Each of them hosts a 3-cm refractor and an uncooled CCD as a detector. The mission objectives and technical details were presented by \cite{2014PASP..126..573W} and \cite{2016PASP..128l5001P}. Due to the limitations of telemetry, only small pieces (subrasters) of the original CCD frames around selected stars are downloaded. As a remedy for the increasing number of hot pixels, occurring as a result of defects caused by cosmic-ray protons, the so-called `chopping mode' was implemented in mid-2015 as a standard way of observing for all BRITEs \citep{2016PASP..128l5001P}. This change resulted in improved photometry, which is performed for observations made in both modes by means of the photometric pipelines described by \cite{2017A&A...605A..26P}. 

\section{Characteristics of the BRITE data}
\begin{figure}[!h]
\centering
\includegraphics[width=0.9\textwidth]{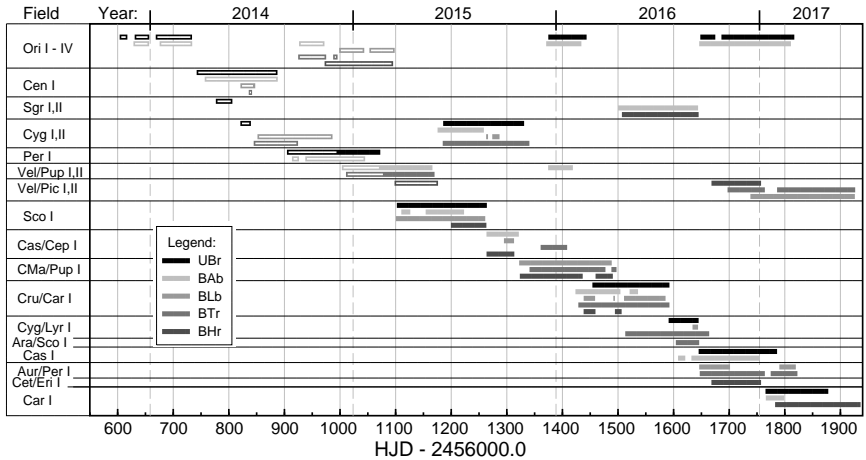}
\caption{Distribution of the observations in BRITE fields completed by October 2017. The data obtained in the stare and chopping observing modes are shown with unfilled and filled bars, respectively. The figure is a modified version of Fig.\,A.1 from \cite{2017A&A...605A..26P}.}
\label{fig:fields}
\end{figure}

Observations performed by BRITEs are organized into runs of up to six months long. During a single run, one to five (typically two or three) satellites observe the same field and stars, selected from the list of potential targets. The run is usually split into several setups\footnote{The concept of a setup is explained by \cite{2017A&A...605A..26P}.}, which mark some changes in the observing conditions, for instance, the change of a subraster position in the detector. In some cases, a given satellite observes two different fields during a single orbit. By the time of writing (October 2017), observations for 24 fields were completed; for 21 of them the photometric data have been released. The names of the BRITE fields follow the name(s) of the constellation(s) in which the observations are conducted; the (Roman) sequential numbers indicate if a field was re-observed. For example, Ori III field means that this is the third run in the Orion field. Figure \ref{fig:fields} shows the distribution of BRITE observations in time for the completed 24 fields.
\begin{figure}[!h]
\centering
\includegraphics[width=0.9\textwidth]{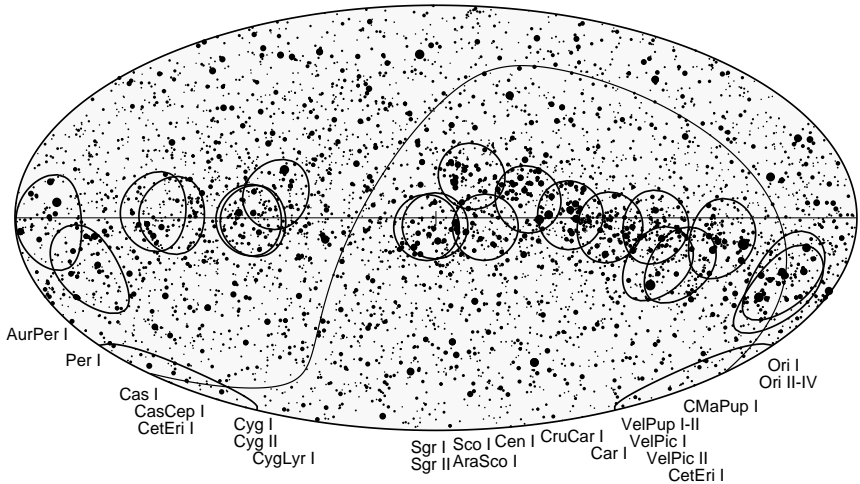}
\caption{Stars brighter than $V=$~6~mag in Galactic coordinates shown in the Aitoff projection. The horizontal line marks the Galactic plane; the curved one, the celestial equator. Twenty four BRITE fields discussed in this paper are encircled and labeled below the map.}
\label{fig:fields-ait}
\end{figure}

The BRITE fields are selected for observations at least six months ahead of their onset to allow planning of the ground-based support.  In principle, BRITEs can observe the whole sky. However, as can be seen in Fig.\,\ref{fig:fields-ait}, the up-to-date observations focused mostly on the fields located along the Galactic plane, especially on the Scorpius-Centaurus complex of nearby associations, the closest part of the Gould Belt. The only exception was the CetEri~I field located relatively far (40$^{\rm o}$\,--\,60$^{\rm o}$) from the Galactic plane. The main reason for such selection is that the fields close to the Galactic plane contain the largest number of scientifically interesting targets. They also contain a large number of bright stars, which are required by the star trackers to achieve the  fine pointing. Some attempts to observe other fields far from the Galactic plane failed because of the lack of bright stars  suitable for tracking. Figure \ref{fig:fields-ait} shows also that southern part of the sky is favoured. In fact, about 61\% of BRITE targets are located in the southern hemisphere. This fact has consequences for the planning of ground-based support for BRITE targets.

The statistics presented in this paper are confined to data and stars for which the photometry has been released, that is, the data for 21 fields, or all 24 fields with completed observations except for AraSco~I, VelPic~II, and Car I.\footnote{The data for the last two fields have now been reviewed; those for AraSco~I will be released together with the second run in the AraSco field.} In total, 426 stars have been observed in these 21 fields, some of them during more than a single run. These observations comprise 34\,$\times$\,10$^6$ individual measurements.

The exposure times for BRITE observations range between 0.1 and 7.5~s, but most of the observations were obtained with the exposure time of 1~s. The exposures are typically separated by a 20\,--\,25-s gap. Typically, a satellite makes observations during about 15 minutes of the $\sim$100-min orbit, which translates into about 50 measurements per orbit on average. Given the short exposures and orbital gaps, the final duty cycle of the satellites is rather low and amounts to 0.36\% for BAb, 0.49\% for UBr, 0.55\% for BLb, 2.46\% for BTr, and 0.69\% for BHr. The higher duty cycle for the two red-filter satellites, BTr and BHr, is a consequence of the more frequent observing with longer exposure times (up to 7.5~s by BTr and up to 5~s by BHr). These longer-exposure observations allow for including fainter scientifically interesting stars in the BRITE science program.

\section{Characteristics of the observed stars}
The number of stars observed by BRITEs during a single run depends on several factors of which the number of bright scientifically interesting targets in the field and the limitations of telemetry are the most important. The latter is usually a bottleneck and limits the number of selected targets. Chopping mode requires larger subrasters, which could be achieved only by the price of a smaller number of observed stars. The average number of stars observed by BRITEs in a single field was 27, but this number ranged from 12 stars in the VelPup\,II field (observed only by BAb) up to 45 stars in the CruCar\,I field (observed by all five satellites). The largest number of stars observed by a {\sl single} satellite in one field is 36 (BTr run in the VelPup\,I field).

\begin{figure}[!h]
\centering
\includegraphics[width=\textwidth]{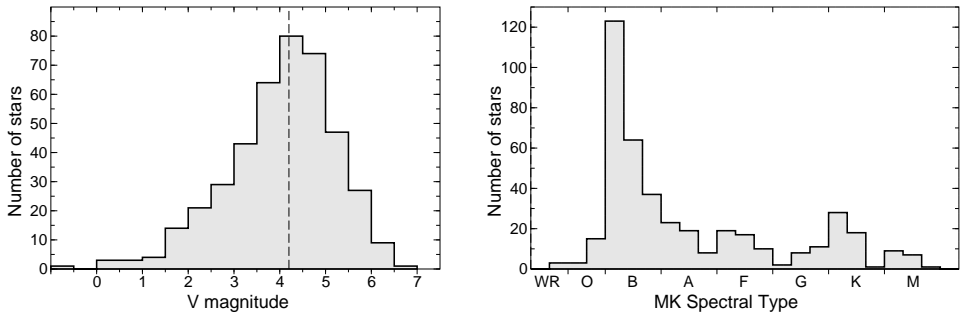}
\caption{Left: Histogram of the $V$ magnitudes of the sample of 426 BRITE targets analyzed in this paper. The vertical dashed line marks the median magnitude ($V=$\,4.20~mag). Right: Histogram of the MK spectral types for the same sample of stars.}
\label{fig:hist-v-spt}
\end{figure}

Prior to the launch of the first BRITE satellites, it was expected that they would observe stars brighter than $V$ magnitude 3.5\,--\,4 \citep{2008CoAst.157..271W,2009AcAau..65..643D,
2014PASP..126..573W}. This magnitude limit would restrict the number of potential BRITE targets to only about 500 of the brightest stars in the sky. Fortunately, it turned out that BRITEs can provide precise photometry for much fainter stars. The statistics of the 426 stars observed by BRITEs shows that the satellites can successfully observe stars much fainter than $V=$~4 mag. In fact (Fig.\,\ref{fig:hist-v-spt}, left panel), the median value of the up-to-now observed targets is 4.20~mag. In other words, more than half of BRITE targets are fainter than 4th magnitude in $V$. By now, the faintest star observed by BRITEs is the 7th magnitude Wolf-Rayet star HD\,50896 = EZ~CMa = WR\,6 located in the CMaPup\,I BRITE field; the brightest star observed so far is Canopus ($\alpha$~Car), the second brightest star in the sky ($V=-$0.7\,mag).

The distribution of spectral types of the observed stars is shown in Fig.\,\ref{fig:hist-v-spt}, right panel. In defining the mission objectives \citep{2014PASP..126..573W}, it was assumed that BRITEs would observe the brightest, but also the most luminous stars in the sky. This is indeed the case: 58\% of the sample are Wolf-Rayet, O and B-type stars at the main sequence and post-main-sequence stages of evolution. This sample is supplemented by a moderate number of A, F, and G-type stars (27\% of the sample), mostly main-sequence objects, but also about a dozen bright Cepheids, and some K and M-type stars (15\%), mostly giants. In general, it can be concluded that BRITE's niche is the study of bright hot luminous stars. By now, BRITEs have gathered observations for 70\% of all O and B-type stars brighter than 4th magnitude in the whole sky.

\section{Science with BRITE data}\label{science}
Precise photometry of bright stars is extremely difficult to obtain from the ground. The lack of good photometry hindered comprehensive studies of many bright objects, especially their variability. Many bright stars usually have a rich record of archival data of other kinds, mostly spectroscopy, but also interferometry and polarimetry. Even if this is not the case, new data of high-quality, especially spectroscopy, can be easily obtained for bright stars using small telescopes. The situation is therefore much better than for many {\it Kepler} targets, which could be accessed spectroscopically only with large telescopes. We have presently about 20 cooperating professional teams who operate ground-based facilities\footnote{See http://www.univie.ac.at/brite-constellation/gbot/gbot-facilities/}, organized via the Ground-Based Observing Team (GBOT; Zwintz, these proceedings) and the developing cooperation with AAVSO amateurs (Kafka, these proceedings). All this means that, in most studies based on BRITE photometry, different complementary data are used.

In addition to the presentation of rough statistics of the data obtained by BRITEs, this paper is also supposed to present considerations on the possible science. We do this referring to different types of objects, focusing on their variability, spectral type, evolutionary stage or other characteristics.

\subsection{Wolf-Rayet stars}
One of the most interesting problems related to Wolf-Rayet (W-R) stars is the origin of their variability (stellar surface and/or stellar wind inhomogeneities). Three W-R stars were observed from space by MOST revealing non-coherent variability interpreted as stochastic. By now, BRITEs observed six W-R stars\footnote{For four (WR 6 = EZ CMa, WR\,11 = $\gamma^2$~Vel, WR\,24, and WR\,48  = $\theta$~Mus) the data have been released; for two more (WR\,24 and WR\,40 = V385~Car), the data will be released soon.}, including the brightest one, $\gamma^2$~Vel, the less massive component of the eccentric W-R + O binary, and a perfect object for the study of interacting winds \citep{2017MNRAS.471.2715R}. The study of BRITE W-R stars should therefore allow for more general conclusions as to the origin of the photometric variability in these stars, because of much better statistics; see also Moffat et al., these proceedings.

\subsection{O-type stars and $\eta$~Car}
For O-type stars, there is a known disagreement between the theory, which predicts pulsations in $p$ and $g$ modes in the full range of spectral type O, and observations, which seem to show that pulsations are excited only in the late O-type stars. This is the conclusion drawn from the ground-based surveys and MOST and CoRoT observations from space. In this regard, BRITEs will significantly increase the sample of surveyed O-type stars.  By now, 17 O-type stars have been observed. The sample includes very massive O4\,I(n)fp star $\zeta$~Pup \citep{2018MNRAS.473.5532R}, three eclipsing stars, UW\,CMa (see Pablo et al., these proceedings), $\delta$~Ori, and $\tau$~CMa, and the most massive known heartbeat star $\iota$~Ori \citep{2017MNRAS.467.2494P}. In addition to pulsations, BRITE data will be useful in deciphering the variability related to mass loss and stellar wind inhomogeneities, in particular in explaining the origin of discrete absorption components (DACs) occurring in stellar line profiles. Examples of such studies are presented by \cite{2017A&A...602A..91B} and \cite{2018MNRAS.473.5532R}. Finally, BRITE observed the famous luminous blue variable (LBV) $\eta$~Car. BRITE photometry should help to understand the nature of its long-term variability (Richardson et al., these proceedings).

\subsection{Be stars}
Be stars are non-supergiant main-sequence B-type stars recognized by their near-critical rotation and the presence of H$\alpha$ line in emission originating in a decretion circumstellar disk. One of the most important problems related to these stars showing variability on time scales ranging from hours to decades is the question how the matter is transferred from the star to the disk and coherent explanation of their spectroscopic and photometric variabilities on all time scales. Spectroscopy provided the first hints that pulsations may play an important role in the process of feeding matter into the disk, but observational evidence was rather poor. Fortunately, the sample of BRITE Be stars is quite rich: 43 objects, i.e.~about 20\% of all B-type stars. Many of them have a very rich spectroscopic record. Combining BRITE photometry, suitable for the studies of the shortest variability scales, and the expertise of the researchers seems to have resulted in a real breakthrough in this topic: there is an increasing evidence that pulsations play a key role in this process \citep{2016A&A...588A..56B,2017arXiv170808413B}. 

\subsection{Main sequence $p$-mode pulsators ($\beta$~Cephei and $\delta$~Scuti-type stars)}
The $\beta$~Cephei stars, $p$-mode pulsators located in the upper main sequence (MS), are potentially very useful for asteroseismology for at least three reasons. First, their seismic modeling may help to reveal details of high-mass star evolution, in particular the extent of the overshooting from the core. Next, many of them rotate relatively rapidly, which may help to understand pulsations in the presence of fast rotation and the evolution of the internal rotation profile at the MS stage. Finally, due to the sensitivity of the driving to the details of the input physics, opacities in particular, pulsations in B-type stars may provide a hint to their revision.  All of this would be possible only if many objects with a relatively large number of detected modes are known.  Up to now, only a handful of such objects have been observed or discovered by other space missions (MOST, CoRoT, WIRE, {\it Kepler}). The situation may change significantly after BRITE.  By now, BRITEs have observed 24 `classical' $\beta$~Cephei stars listed by \cite{2005ApJS..158..193S}, but the list of potential stars of this type is much longer --- about 120 stars in the $\beta$~Cep spectral type range (B0 -- B2.5). Bearing in mind that the detection threshold in BRITE data for some stars is as low as 0.15~mmag, we can expect that BRITE photometry will enable to discover many new modes in known stars as well as new stars of this type. This expectation has been met.  A few examples: in $\beta$~Cen 19 modes were detected in BRITE data \citep{2016A&A...588A..55P}, while only two were known earlier; in $\sigma$~Sco (Pigulski et al., these proceedings), these numbers are 10 vs.~4; for $\theta$~Oph, 16 vs.~7, for $\kappa$~Sco, 14 vs.~4 (Walczak et al., in prep.). Even for $\nu$~Eri, a target of several ground-based campaigns, BRITE data allowed to discover new $g$ modes \citep{2017MNRAS.464.2249H}. A general picture that arises from these findings is that $p$/$g$-mode hybridity is very common among $\beta$~Cep stars, but usually can be found only if amplitudes below 1~mmag are detected. 

The other $p$-mode pulsators at the MS are $\delta$~Scuti-type stars of spectral types ranging from early A to mid-F. There are about 70 stars observed by BRITEs falling in this spectral range. This sample includes some known or even famous $\delta$~Sct stars like $\beta$~Pic (Zwintz et al., these proceedings), but we expect to discover new ones as well.

\subsection{Main sequence $g$-mode pulsators (SPB and $\gamma$~Doradus-type stars)}
About 200 stars observed by BRITEs are potential slowly pulsating B-type (SPB) stars. Due to the high density of $g$ modes in B-type star models, their identification is difficult and relies usually on searching for regular patterns, the result of the asymptotic relations obeyed by periods of the $g$ modes. Some interesting results on individual SPB stars were already obtained from BRITE data. In particular, \cite{2017A&A...603A..13K} found rotationally split modes in the frequency spectrum of the late B-type SPB star V369\,Cyg and concluded that the core of the star rotates much more slowly than the envelope, which due to tidal effects synchronizing its rotation with the orbital period of the binary. A review of the pulsational characteristics of a large sample of BRITE SPB type stars is now in preparation (Niemczura et al., priv.~comm.) One of the interesting findings of BRITE related to SPB stars is the detection of many stars with only $g$ modes among early B-type stars \citep[][and these proceedings]{InnsbruckJDD}, that is, in the spectral region in which theory predicts the instability of both $p$ and $g$ modes. For some reason, only the latter reach detectable amplitudes.

The other group of $g$-mode main-sequence pulsators on the MS are $\gamma$~Dor stars having spectral types F. There are not many stars of this spectral type in the BRITE sample (Fig.\,\ref{fig:hist-v-spt}), but still about 20 stars can be regarded as potential $\gamma$~Dor pulsators. One of them, 43 Cyg, is a known star of this type. BRITE data allowed to find at least 43 $g$ modes in this star and identify the period spacing pattern \citep[][van Reeth et al., these proceedings]{2017A&A...608A.103Z}. This, in turn, allowed to put some constraints on the core rotation and even derive the near-core rotation rate.

\subsection{B- and A-type supergiants}
The variability of B and A-type supergiants remains poorly understood. Several were observed by MOST and CoRoT, but one month long observations seemed to be too short to resolve complicated variability in these stars. Therefore, 14 B-type and four A-type supergiants in the BRITE sample with up to six-month long observing runs are ideal for the study of variability in these stars. Some preliminary work has been done (Guinan et al., in prep.; Rybicka et al., these proceedings), but in general the BRITE data for hot supergiants are not yet well exploited. These stars are extremely interesting in the context of the theoretically predicted pulsations, driving mechanism(s), the presence of strange modes and stochastic oscillations. These studies can allow to estimate mass-loss rates and evolutionary status of these stars.

\subsection{Magnetic B-type stars}
It is extremely interesting that magnetic fields are discovered in stars of practically all spectral types, including B-type stars having radiative envelopes. There is a BRITE-related project, the BRITE Spectropolarimetric Survey \citep{InnsbruckNeiner}, a magnitude-limited ($V<4$ mag) survey aimed at the detection of magnetic fields. One of the first results of this survey is finding that about 10\% of stars show detectable magnetic fields. Many stars covered by the survey were observed by BRITEs and the first interesting results from the combination of BRITE photometry and spectropolarimetric observations were obtained \citep[][Buysschaert et al., these proceedings]{InnsbruckWade}. The most interesting scientific problems related to magnetic fields in hot stars are the relations between pulsations and magnetic fields, and the incidence and topology of magnetic fields.

\subsection{Red giants and Cepheids}
There are about 55 red giants in the BRITE sample. This is not a small sample, but in comparison with the samples observed by CoRoT and {\it Kepler}, especially in the terms of photometric quality and time coverage (for {\it Kepler}), one cannot expect to get outstanding results from BRITE data. To date, no convincing detection of solar-like oscillations was found in BRITE data. One of the possible applications of BRITE data to red-giant science is the possibility of empirical calibration of scaling relations provided that radii of the stars could be determined empirically (Hekker \& Huber, unpublished Sobieszewo presentation). This would be possible for stars having stellar diameters determined from the interferometric measurements of angular diameter and a knowledge of the distance. Whether this is possible for any of the BRITE targets remains to be shown.

By now, 11 bright Cepheids have been observed by BRITEs, increasing considerably the number of Cepheids observed from space. The preliminary results for some of them were published by \cite{InnsbruckSmolec}. The study is focused on searches for the Blazhko effect, period doubling, and non-radial modes. Some evidence for the occurrence of some of these effects in the studied stars was found.

\subsection{Binaries}
Binarity is very common especially among early-type stars. Consequently, a majority of BRITE targets belong to binaries. Of 426 stars in the BRITE sample discussed in this paper, at least 202 (47\%) are known members of spectroscopic binaries (SBs), at least 256 (60\%) are members of visual binaries (VBs), and at least 32 (8\%) are eclipsing binaries (EBs). These are lower limits, since many SBs or EBs were discovered from BRITE or BRITE-associated ground-based data (for example, Pigulski et al., these proceedings). In general, binarity is a big advantage because in many cases it allows to derive fundamental stellar parameters (masses and radii), crucial in other astrophysical applications. Examples include seismic modelling \citep[for example,][]{2016A&A...588A..55P} or tests of the theory of stellar evolution. In addition, a wealth of astrophysical phenomena can be studied in binaries: tidal interactions and tidally excited oscillations \citep[][Pigulski et al., these proceedings]{2017MNRAS.467.2494P}, mass exchange (Rucinski et al., these proceedings), and stellar wind collisions \citep{2017MNRAS.471.2715R}.

Concluding, one can ask the question: Is there any missing science that can be done with BRITE data? It is always difficult to answer such a question. Given the mission objectives \citep{2014PASP..126..573W} and the expectations \citep{2012arXiv1211.5439W}, it seems that BRITE data should be checked for the possible transits of exoplanets. There is no other obvious area in which BRITE data may contribute, but the data studied by now show that some serendipitous and valuable discoveries can be expected.

\acknowledgements{The study is based on data collected by the BRITE-Constellation satellite mission, designed, built, launched, operated and supported by the Austrian Research Promotion Agency (FFG), the University of Vienna, the Technical University of Graz, the Canadian Space Agency (CSA), the University of Toronto Institute for Aerospace Studies (UTIAS), the Foundation for Polish Science \& Technology (FNiTP MNiSW), and National Science Centre (NCN). The operation of the Polish BRITE satellites is secured by a SPUB grant of the Polish Ministry of Science and Higher Education (MNiSW). The author acknowledges support from the NCN grant no.~2016/21/B/ST9/01126.
}

\end{document}